\documentclass[amsfonts,showpacs,tightenlines,aps,12pt,floatfix]{revtex4}
\usepackage{bm}
\begin{document}

%\preprint{}

\title{Heavy Quark Analogues of the $\Theta$ and Their Excitations}
\author{Kim Maltman}
\email[]{kmaltman@yorku.ca}
%\thanks{}
\affiliation{Department of Mathematics and Statistics, York University, 
4700 Keele
St., Toronto, ON CANADA M3J 1P3}
\altaffiliation{CSSM, Univ. of Adelaide, Adelaide, SA 5005 AUSTRALIA}

\date{\today}

\begin{abstract}
Predictions for the low-lying excitation spectrum of positive parity
pentaquark systems containing one $\bar{c}$ or $\bar{b}$ 
antiquark and four light $u,d$ quarks are obtained in the quark model
picture for models with spin-dependent interactions given either by 
effective color magnetic (CM) exchange or effective Goldstone boson 
(GB) exchange. 
For the GB model, 4 excited states 
are predicted to lie within $\simeq m_\Delta -m_N$ of the $J^P=1/2^+$ 
ground state while, for the CM model, 10 states are
expected in the same range. Both the lowest excitation energy
and the relative splittings are much smaller in the CM case.
These predictions are on the same footing 
as those for the analogous splittings in the
non-exotic baryon sector and, as such, provide
a means of not only testing the models, but potentially ruling out 
either one, or both.
\end{abstract}

\pacs{12.39.Mk,14.20.Lq,14.20.Mr,12.40.Yx}
%\keywords{}

\maketitle

\section{\label{intro}Introduction}
Interest in ``heavy pentaquarks'' (states with the quark content
$\bar{Q}q^4$, where $\bar{Q}=\bar{c}$ or $\bar{b}$, $q=\ell$ or
$s$, with $\ell =u$ or $d$) goes back more than fifteen years
to observations made in the context of the effective color-magnetic
exchange (CM) model, a model with considerable phenomenological success in
describing splittings in the baryon spectrum. The spin-dependent
interactions of the model have the form
\begin{equation}
H_{CM}=\sum_{i<j}{\frac{C_{CM}}{m_im_j}}\, f_{CM}
(\vec{r}_{ij})\, \vec{\sigma}_i\cdot\vec{\sigma}_j\, 
\vec{F}_i\cdot\vec{F}_j\ ,
\label{hhfcm}\end{equation}
where $\vec{\sigma}_i$ are the Pauli spin matrices,
$\vec{F}_i=\vec{\lambda}_i/2$ for quarks and 
$\vec{F}_i=-\vec{\lambda}^*_i/2$ for antiquarks, are the 
Gell-Mann color matrices, $C_{CM}$ is a constant,
$m_i$ is the constituent quark mass, and $f_{CM}(\vec{r}_{ij})$ contains
the spatial dependence, usually taken to be a smeared version
of the delta function. In the combined $SU(3)_F$ and 
$m_{\bar{Q}}\rightarrow\infty$ limits, the $J^P=1/2^-$, flavor $3_F$ 
heavy pentaquark channel was found to have a hyperfine
expectation optimally attractive relative to 
the corresponding strong decay threshold, 
$BP_H$ (with $B$ and $P_H$ the relevant octet baryon and heavy 
pseudoscalar meson)~\cite{earlyheavy}.
Subsequent investigations showed that $SU(3)_F$ breaking,
kinetic energy, confinement, and 
$m_{\bar{Q}}\not= \infty$ effects all reduced
binding. The combined effect made it unlikely
that even the most favorable of the 
$J^P=1/2^-$, $3_F$ states (that with isospin and strangeness 
$(I,S)=(1/2,-1)$) would bind~\cite{nextheavy}.
Since an attractive s-wave interaction 
insufficiently strong to bind produces positive phase motion, 
but not resonant behavior, such a $J^P=1/2^-$ state, if above threshold,
would be non-resonant, with 
an s-wave ``fall-apart''decay to $ND_s$ or $NB_s$.

The situation is rather different for the Goldstone boson
exchange (GB) model, where negative parity heavy pentaquark 
states were found to be unbound by several $100$ MeV~\cite{gbnegparheavy}. 
The model involves effective interactions generated
by Goldstone boson exchange, and was introduced to deal with certain
phenomenological problems of the CM model, in particular
the problem of the incorrect ordering of 
positive and negative parity excited baryon states~\cite{glozman96}. 
The effective spin-dependent GB $qq$ interactions have the form
\begin{equation}
H_{GB}=\sum_{i<j}{\frac{C_{GB}}{m_im_j}}
f_{GB}(\vec{r}_{ij})\, \vec{\sigma}_i\cdot\vec{\sigma}_j\, 
\vec{F}_i\cdot\vec{F}_j\ ,
\label{hhfgb}\end{equation}
with $\vec{\sigma}_i$ the Pauli spin matrices, 
$\vec{F}_i=\vec{\lambda}_i$ the Gell-Mann $SU(3)_F$
flavor matrices, and $C_{GB}$ a constant. The explicit form of 
$f_{GB}(\vec{r}_{ij})$, whose parameters
are fixed from the study of the baryon spectrum,
is given in Ref.~\cite{gbposparheavy}. As first noted by 
Stancu~\cite{gbposparheavy}, reducing the 
$q^4$ orbital permutation symmetry from $[4]_L$ to $[31]_L$ by 
introducing a single p-wave excitation among the four light quarks
increases the hyperfine attraction. The interaction turns out to bind 
those positive parity ($P=+$), 
$6_F$ states having quark content $\bar{Q}\ell^4$
and discrete quantum numbers $(I,S,J_q)=(0,0,1/2)$ (where $J_q$ is 
the total intrinsic spin which, combined with the orbital $L=1$, 
yields the total $J$). The binding energies
are $76$ MeV and $96$ MeV for the $\bar{Q}=\bar{c}$ and $\bar{b}$ 
systems, respectively~\cite{gbposparheavy}.
The other members of the $6_F$ multiplet are predicted to be unbound,
but by less than $\sim 100$ MeV, and hence might appear as genuine
resonances, since an attractive interaction in a relative
p-wave can play off against the peripheral centrifugal repulsion
to produce resonant behavior.

An experimental search for strong interaction stable
anticharmed $(I,S)=(1/2,-1)$ pentaquark states decaying
to $K^{*0}K^-p$ was performed by the E791 Collaboration~\cite{e791heavy}. 
Negative results were reported in the mass range $2.75-2.91$ GeV.

The discovery of the $S=+1$ $\theta$ baryon~\cite{thetaexp}
has led to a revived interest in heavy
pentaquarks~\cite{H1,zeus,jw,klth,huang,rosner,sww,csmthetac,otherqm}. 
If, as is now generally assumed,
the $\theta$ parity is positive, then whatever
mechanism makes the $\theta$ narrow is likely to also 
make its heavy quark analogues, the $\theta_{c,b}$, narrow.
%, 
%even if they lie above the $N\bar{D}$ and $NB$ strong decay thresholds.
The H1 collaboration has recently presented evidence for
a narrow anticharmed pentaquark resonance, with mass $3099\pm 3\pm 5$ MeV
and width compatible with experimental resolution, decaying to 
$D^{*-}p$~\cite{H1}. This observation has yet to be confirmed.
The H1 state was also searched for, but not seen, by ZEUS~\cite{zeus}.

A number of recent quark-model estimates exist 
for $m_{\theta_{c,b}}$. These are 
based on (i) existing proposals for the structure of the $\theta$,
(ii) assumptions about the relation between the 
structures in the $\theta$ and $\theta_{c,b}$
and (iii) the experimental value 
$m_\theta \simeq 1540$ MeV~\cite{jw,klth,huang}.

In the Jaffe-Wilczek (JW) scenario~\cite{jw,nussinov}, a structure 
consisting of two $I=J=0$, $C=\bar{3}$ $qq$ pairs coupled
antisymmetrically to net color $3$, with only confinement
forces between the $\bar{s}$ and the $qq$ pairs,
is proposed for the $\theta$. The same light quark configuration 
is then expected for the $\theta_{c,b}$.
The analogous baryon splittings,
$\simeq m_{\Lambda_{c,b}}-m_\Lambda$,
are used to estimate $m_{\theta_{c,b}}-m_\theta$.
The resulting estimates,
\begin{eqnarray}
&&\left[ m_{\theta_c}\right]_{JW}\simeq 2710\ {\rm MeV}\nonumber\\
&&\left[ m_{\theta_b}\right]_{JW}\simeq 6050\ {\rm MeV}\ ,
\label{jwthetac}\end{eqnarray} 
lie $\sim 100$ and $170$ MeV, respectively, below 
$N\bar{D}$ and $NB$ thresholds.

In the Karliner-Lipkin (KL) scenario, the structure proposed
for the $\theta$ consists of one
$I=J=0$, $C=\bar{3}$ $qq$ pair, as in the JW scenario, but
with the spin and color of the remaining pair flipped
and anti-aligned to those of the $\bar{s}$, producing ``triquark'' 
($ud\bar{s}$) quantum numbers $(I,J,C)=(0,{\frac{1}{2}},3)$~\cite{kl}.
The scenario is motivated by the 
CM model, in which the hyperfine energy of the KL correlation
is lower than that of the JW correlation. The $q\bar{Q}$ hyperfine 
interactions (which drive the $ud\bar{s}$ cluster formation) 
are weakened when $\bar{s}$ is replaced by $\bar{c},\bar{b}$,
reducing the hyperfine attraction in the $\theta_{c,b}$ systems.
KL estimate this reduction by
assuming that (i) the same diquark-triquark correlation is
present in the $\theta_{c,b}$ as in the $\theta$ and (ii) the
strength of the $\bar{Q}$ hyperfine interactions scale, as in
the CM model, with the inverse of the constituent $\bar{Q}$
mass~\cite{klth}. The resulting modification of the
JW estimates leads to
\begin{eqnarray}
&&\left[ m_{\theta_c}\right]_{JW}\simeq 2985\ {\rm MeV}\nonumber\\
&&\left[ m_{\theta_b}\right]_{JW}\simeq 6400\ {\rm MeV}\ ,
\label{klthetac}\end{eqnarray} 
which puts the $\theta_c$ and $\theta_b$ both
$\simeq 180$ MeV above strong decay threshold.

The JW and KL estimates for $m_{\theta_{c,b}}$
rely on the assumption that the change in $1$-body energies
in going from the $\theta$ to $\theta_{c,b}$ is well approximated
by the analogous change in the $q^3$ sector. The systematic
uncertainty accompanying this assumption is difficult to estimate. 
One should bear in mind that, in contrast to
the chiral soliton model picture, where a low-lying $\theta$ is quite 
natural~\cite{earlycsm,dpp,csmothers,ekp}, the $\theta$ mass is much lower than
naive constituent quark model expectations would have anticipated.
Such expectations are, however, based on effective
quark model Hamiltonian which do not explicitly take into
account differences in vacuum response in the $q^3$ and pentaquark sectors. 

The dibaryon sector of the bag model provides an
illustration of the problems such neglect might produce~\cite{kmdib}.
The difference between the $1$-body energy of a single $q^6$
bag and that of two isolated $q^3$ bags, evaluated
with standard bag model parameters, is $\sim 50$ MeV. This
relatively small shift, however, results from a 
close cancellation between two $\sim 400$ MeV shifts, one in
the kinetic, and one in the ``zero point'' ($Z/R$) energy. This cancellation
is an extremely sensitive function
of the bag parameter, $B^{1/4}$~\cite{kmdib}. 
Sizeable uncertainties are thus present in estimates for the 
$1$-body contribution to splittings between ordinary hadron and
multiquark states, even before the relative crudeness of the
modelling of the vacuum response in the bag model
is taken into account. Typical constituent
quark models for which extensions to the $P=+$
pentaquark sector are feasible do not explicitly incorporate
even such a simple realization of vacuum response.
The resulting estimates for $1$-body contributions to pentaquark energies 
are thus likely to have sizeable uncertainties.
Predictions insensitive to the model treatments of the $1$-body energies
are thus desirable. We discuss several such predictions in this paper.

The fact that the $\theta$, and hence its
partners in the $\overline{10}_F$ multiplet having $N$ and
$\Sigma$ quantum numbers, lie in the midst of the
first positive parity excitation baryon region also suggests
that past treatments of the excited baryon sector which
include only $q^3$ configurations and neglect mixing with pentaquark
states are almost certainly unreliable. The phenomenological
successes (or failures) of the models in accounting for the excited baryon
spectrum thus need to be revisited. Since the presence of
pentaquark configurations makes the phenomenology of the
excited baryon sector more complicated than heretofore
anticipated, it is useful to have distinctive predictions
of the models in the phenomenologically less-complicated exotic
sector. The results of this paper, which show significant differences
for the {\it splittings} in the $P=+$ heavy pentaquark sectors
of the GB and CM models, provide useful predictions of this type.

In the rest of the paper, we present the results of 
fully-antisymmetrized GB and CM model calculations
for the hyperfine energies of heavy $P=+$, $\bar{Q}\ell^4$
states. The results allow us to investigate,
in a dynamical context, cross-cluster interaction
and antisymmetrization effects neglected in the JW and KL approaches,
and to study the impact of such effects on the
JW and KL estimates for the heavy pentaquark hyperfine energies.
These points are discussed in Sec.~\ref{IIA}. In Sec.~\ref{IIB}
we discuss the splittings predicted by 
each model. These are determined up to an overall
scale, associated with the size of the $[31]_L$
spatial wavefunction, by the spin-isospin-color structure
of the effective interactions, and are
independent of the model $1$-body energies.
This would not be true of the splittings
between $\bar{Q}\ell^4$ and $\bar{Q}s^n\ell^{4-n}$ states, 
where flavor-breaking effects for the
problematic $1$-body energies would need to be taken into account. 
In Sec.~\ref{IIB} we also present results for the 
overlaps of the various pentaquark states to
$NP_H$ and $NV^*_H$ (with $V^*_H$ the relevant heavy
vector meson). Ratios of these overlaps are expected to determine
the ratios of effective couplings to the $NP_H$ and $NV^*_H$ decay 
channels, if the decay mechanism is dominated by
``fall-apart'' through the p-wave centrifugal barrier~\cite{closezhao}.

\section{\label{section2}Heavy Pentaquark States in the GB and CM Models}
The results which follow are obtained by constructing, in each
$(I,J_q)$ channel, all possible color singlet pentaquark
states obtainable from fully-antisymmetrized $\ell^4$
states with $[31]_L$ orbital and $[211]_c$ color symmetry,
and diagonalizing $H_{GB}$ or $H_{CM}$ in the resulting basis.
The construction of states and evaluation of matrix elements are
standard, and not presented here. Useful cross-checks on the state
construction, phase conventions, and spin, color and isospin
matrix elements employed are provided 
by the schematic versions of $H_{GB,CM}$, which neglect 
the spatial dependence of the operators, allowing
the expectations to be obtained by standard group theoretic methods. 

In the spatial sector, it is convenient to work with a 
$[31]_L$ $S_4$ basis whose members transform
as the $SS$, $SA$ and $AS$ irreps of $S_2^{12}\times S_2^{34}$.
The GB and CM light quark hyperfine matrix elements 
are then completely determined by the spatial matrix elements 
\begin{equation}
\langle [31]_L,\rho\vert f_{GB,CM}(\vec{r}_{12})\vert [31]_L,\rho\rangle ,\ 
\rho =SS,SA,AS
\end{equation}
To estimate the contributions from the $\bar{Q}\ell$ 
interactions, which vanish identically only 
in the $m_{\bar{Q}}\rightarrow\infty$ limit, we
employ the following form for the $\rho =SS,SA,AS$ $[31]_L$
spatial wavefunctions (where only the $(L,L_z)=(1,0)$
component is displayed):
\begin{equation}
\psi_{10}^\rho \left(\vec{r}_{SS},\vec{r}_{SA},\vec{r}_{AS},\vec{R}_5\right)\,
 =\, N\, z_\rho\, exp\left[ -{\frac{\alpha^2}{2}}\left( r_{SS}^2+r_{SA}^2+
r_{AS}^2\right) -{\frac{\beta^2}{2}}R_5^2\right]\ .
\label{spatialwf}\end{equation}
$N$ is a normalization constant, and 
the relative coordinates,
$\vec{r}_{SS}$, $\vec{r}_{SA}$, $\vec{r}_{AS}$, and $\vec{R}_5$,
are defined in terms of the quark ($\vec{r}_i,\ i=1,\cdots ,4$)
and antiquark ($\vec{r}_5$) coordinates, by
$\vec{R}_5=\sqrt{{\frac{4}{5}}}\left[{\frac{1}{4}}\left(
\vec{r}_1+\vec{r}_2+\vec{r}_3+\vec{r}_4\right) -\vec{r}_5\right]$,
$\vec{r}_{SS}={\frac{1}{2}}
\left( \vec{r}_1+\vec{r}_2-\vec{r}_3-\vec{r}_4\right) /2$,
$\vec{r}_{SA}={\frac{1}{\sqrt{2}}}\left( \vec{r}_3-\vec{r}_4\right)$, and
$\vec{r}_{AS}={\frac{1}{\sqrt{2}}}\left( \vec{r}_1-\vec{r}_2\right)$.
%The subscripts $SS$, $SA$, $AS$ give the $S_2^{12}\times S_2^{34}$
%labels. 
%The vectors $\vec{r}_{SS}$, $\vec{r}_{SA}$, and $\vec{r}_{AS}$
%form the basis of a $[31]$ $S_4$ irrep.
We quote results below in dimensionless form by removing a factor of 
$\langle [31]_L,SS\vert f_{GB,CM}(\vec{r}_{12})\vert [31]_L,SS\rangle$
from all hyperfine expectations.

For the GB model, we work with the version 
employed in Refs.~\cite{gbposparheavy,jm},
in which GB exchange is considered
only between quarks, and not between the
light quarks and relevant antiquark, $\bar{Q}$.
The dominant model uncertainty in the predictions for the
splittings and overlaps is associated with the $\alpha$ 
dependence of the ratio, $\mu_{AS}^{GB}$, where
\begin{equation}
\mu_{AS}^{GB,CM}\equiv \langle [31]_L,AS\vert f_{GB,CM}(\vec{r}_{12})
\vert [31]_L,AS\rangle
/\langle [31]_L,SS\vert f_{GB,CM}(\vec{r}_{12})\vert [31]_L,SS\rangle\ .
\label{muasdefn}\end{equation}
For the variationally optimized value 
of $\alpha$ found in Ref.~\cite{gbposparheavy}, $\mu^{GB}_{AS}=0.32$. 
We allow a $\pm 50\%$ variation about this value to study 
the possible model dependence of the results.

In the CM model, the residual $\bar{Q}\ell$ interactions present when
$m_{\bar{Q}}\not= \infty$ lead both to small shifts in the 
$m_{\bar{Q}}\rightarrow\infty$ hyperfine expectations, and to 
additional mixing, in particular between configurations with different light 
quark spin. This can have a non-trivial impact on the overlaps to
$NP_H$ and $NV^*_H$. The size of these effects depends 
on $\beta /\alpha$. We study this dependence
by varying $\beta /\alpha$ between $0.4$ and $0.8$.
The midpoint of this range corresponds to the variational
solution of Ref.~\cite{gbposparheavy}.
% The heavy antiquark is bound to the light quark system
% only by the confining force in Ref.~\cite{gbposparheavy}.
The dominant model dependence for the light quark hyperfine
expectations is that associated with $\mu_{AS}^{CM}$. Taking
a Gaussian form, 
$f_{CM}(\vec{r})={\frac{\sigma^3}{\pi^{3/2}}}\, exp[-\sigma^2 r^2]$, 
for $f_{CM}(\vec{r})$,
$\mu_{AS}^{CM}$ depends on $\alpha /\sigma$. This
ratio is varied between $0$ (zero range limit) and $0.5$
(corresponding to a range of $\sim 1/3\ fm$). $\mu^{CM}_{AS}$,
of course, vanishes in the zero range limit.

\subsection{\label{IIA}GB and CM Model Perspectives on the JW and KL 
Predictions for $m_{\theta_{c,b}}$}
The JW $I=J=0$, $C=\bar{3}$ $qq$ correlation is by far
the most attractive $qq$ correlation in the GB model. 
The version of the GB model employed here, having
no $\bar{Q}\ell$ hyperfine interactions, also matches the
JW scenario for the $\bar{Q}q$ interactions 
in both the $\theta$ and $\theta_{c,b}$
channels, and hence provides a model context in
which the impact of cross-cluster antisymmetrization
and interaction effects, neglected in the JW approach,
can be investigated. Factoring out $C_{GB}/m_\ell^2$
and the $qq$ pair spatial
matrix element, $\langle H_{GB}\rangle =-8$ for each
$I=J=0$ $qq$ pair. Two such pairs are thus hyperfine
attractive relative to the $N$, whose
expectation is $-14$ (corresponding to $\sim -410$ MeV).
Once one accounts for cross-cluster interactions and
antisymmetrization, the model allows admixtures of
other, higher-lying configurations.
%(for example, that
%having two $I=J=1$, $C=\bar{3}$ pairs, for which
%$\langle H_{GB}\rangle =-4/3$ per pair). The net effect of all 
corrections, in the $(I,J_\ell ,J_q )=(0,0,{\frac{1}{2}})$
ground state, is to lower the dimensionless hyperfine
expectation from $-16$ to $-21.9\pm 1.3$. As in the JW
scenario, the light quark configuration is the same for
the $\theta$ and $\theta_{c,b}$. The ground
state turns out to be rather close to a pure
$[4]_{FJ}$ configuration, a result which would be
exact in the schematic approximation. The ground
state expectation, however, differs significantly from the schematic
value, $-28$, as a consequence of the reduction
of $\mu_{AS}^{GB}$ from its schematic limit value, $1$.

Although the $I=J=0$, $C=\bar{3}$ correlation is also the most attractive $qq$
correlation in the CM model, more complicated correlations
yield $\bar{Q}=\bar{s}$ pentaquark configurations with hyperfine 
energies below that of the JW ansatz~\cite{kl}.
The KL diquark-triquark correlation is among these.
However, even lower-lying configurations exist~\cite{jm}. Indeed, for 
the $P=+$ ground state channel, the same 
$\bar{Q}\ell$ interactions which lower the KL $ud\bar{s}$ hyperfine 
expectation also mix the KL and JW configurations.
In the $\theta$ sector, the lowest eigenvalue,
computed using the full set of fully-antisymmetrized 
states, turns out to be reproduced rather accurately (to within $\sim 1\%$)
by a restricted two-channel calculation employing the KL, JW basis
which incorporates this mixing but ignores completely cross-cluster 
antisymmetrization and interaction effects~\cite{jm}. This suggests that, 
for the CM model, the $\theta$ is dominated by the optimized 
combination of JW and KL configurations. 
%This combination turns out to be a roughly equal admixture
%of the two correlations.

The implications of the above discussion for the
$\theta_{c,b}$ follow from the $1/m_{\bar{Q}}$
dependence $\bar{Q}\ell$ interactions in the CM model, which effect
reduces both the KL triquark hyperfine attraction and the
strength of the mixing between JW and KL correlations, relative
to the $\theta$. 
With the conventional constituent quark mass values of Ref.~\cite{kl}, 
the hyperfine attraction is greater for the JW correlation
than the KL correlation, for both the $\theta_c$ and $\theta_b$. 
The KL ansatz, in which the same diquark-triquark correlation posited for 
the $\theta$ is assumed to dominate the $\theta_{c,b}$, 
thus requires additional dynamics beyond that of the CM model.
With strictly CM model interactions, 
in the $m_{\bar{Q}}\rightarrow\infty$ limit,
one expects an $(I,J_q)=(0,1/2)$ ground state in the $P=+$ sector,
dominated by the JW correlation. Neglecting cross-cluster
interaction and antisymmetrization effects, as well as mixing with other 
configurations, the hyperfine expectation for the JW correlation is, after
factoring out $C_{CM}/m_\ell^2$ and the $qq$ pair spatial matrix element, 
$\langle H_{CM}\rangle_{JW} =-4$. The true ground state expectation
in the model, $\langle H_{CM}\rangle =-3.48\pm 0.04$,
obtained from the full $m_{\bar{Q}}\rightarrow\infty$ limit calculation,
indeed occurs for the
$(I,J_\ell ,J_q)=(0,0,1/2)$ channel and is reasonably approximated
by $\langle H_{CM}\rangle_{JW}$. The ground state expectations
for the $\bar{c}$ and $\bar{b}$ pentaquark channels are in turn
well approximated by the $m_{\bar{Q}}\rightarrow\infty$ limit results.

The difference in the hyperfine energies of the $\theta$ and $\theta_{c,b}$
in the CM model can be estimated using the $N$ spatial matrix element 
(which is fixed by the $\Delta$-$N$ splitting) to approximate the 
corresponding $\theta_{c,b}$ spatial matrix elements. With this estimate, 
the reductions in the $\theta_c$, $\theta_b$ hyperfine energies, relative to 
the $\theta$, are found to be $125\pm 30$ and $129\pm 29$ MeV, respectively.
Combining these results with the JW/KL estimates for the $1$-body
contribution to $m_{\theta_{c,b}}-m_\theta$, one obtains the modified estimates
\begin{eqnarray}
&&m_{\theta_c}\simeq 2835\pm 30\ {\rm MeV}\nonumber\\
&&m_{\theta_b}\simeq 6180\pm 30\ {\rm MeV}\ ,
\label{modifiedthetac}\end{eqnarray} 
which put the $\theta_c$ just above and the $\theta_b$ 
just below the corresponding strong decay thresholds.
The errors quoted in Eqs.~(\ref{modifiedthetac})
reflect only uncertainties in the estimated hyperfine
energies. A sizeable additional uncertainty should presumably also
be attributed to the baryon-mass-difference-based
estimate for the $1$-body energy shift between the $\theta$
and $\theta_{c,b}$.
%and the nearness of the
%predictions in Eqs.~(\ref{modifiedthetac}) to threshold,
These uncertainties make a reliable conclusion about 
the strong interaction stability (or instability)
of the $\theta_{c,b}$ in the CM model impossible.

\subsection{\label{IIB}Splittings and Decay Overlaps
in the Heavy Pentaquark Sector}
While uncertainties in the estimates of $1$-body energy shifts mean that
model predictions for the masses of $P=+$ heavy pentaquark 
states are subject to considerable uncertainties, 
the same is {\it not} true of predictions for the splittings
between low-lying excitations and 
the corresponding $(I,J_q)=(0,1/2)$ ground state.
Up to an overall scale, these splittings are determined
by the spin-flavor (or spin-color) structure of $H_{GB}$ (or $H_{CM}$),
and are on the same footing as predictions for the splittings in
the ordinary baryon spectrum. 

Predictions for $P=+$ heavy pentaquark splittings
in the GB and CM models are given in Tables \ref{tablegb}
and \ref{tablecmcb}. Column 1 gives the $(I,J_q)$ quantum numbers
of the states, Column 2
$\Delta\hat{E}$, the hyperfine splitting relative to
the $(I,J_q)=(0,1/2)$ ground state, in units of 
$X=C_{GB,CM}\langle [31]_L,SS\vert 
f_{GB,CM}(\vec{r}_{12})\vert [31]_L,SS\rangle /m_\ell^2$.
$X$ is determined by the $\Delta$-$N$ splitting
in the limit that the pentaquark $SS$ $ij=12$ spatial pair 
expectation is the same as that in the $N$. 
Results for the splittings in this limit, $\Delta E^{est}$,
are given in Column 3. States with $\Delta E^{est}$ greater
than $\sim m_\Delta -m_N$ have been omitted from the tables. Deviations of the 
pentaquark relative-s-wave pair distribution from that in the
$N$ will produce only a global rescaling of all splitting values.
The remaining columns give the squares of the relative overlaps to
$NP_H$ and $NV_H^*$. These
entries are discussed in more detail below. The range given
for each entry reflects the impact of varying
$\beta /\alpha$, $\alpha /\sigma$, and $\mu_{AS}^{GB}$ within
the bounds specified above.

As stressed by Close and Zhao~\cite{closezhao}, if the dominant
mechanism for $P=+$ pentaquark decay to $NM$ ($M=P_H, V_H^*$) 
is ``fall-apart'' through the p-wave barrier, 
the relatives widths for the decays
$P_1\rightarrow NM_1$ and $P_2\rightarrow NM_2$, should be given by
\begin{equation}
{\frac{\Gamma [P_1\rightarrow NM_1]}{\Gamma [P_2\rightarrow NM_2]}}
\, =\, {\frac{\rho_1(m_{P_1})}{\rho_2(m_{P_2})}}\left[
{\frac{\langle NM_1\vert P_1\rangle}{\langle NM_2\vert P_2\rangle}}
\right]^2\ ,
\end{equation}
with $\rho_k(m_{P_k})$ the phase space factor for $P_k\rightarrow NM_k$. 
%The
%discrete part of the overlap was shown to be naturally
%small in the JW, KL and mixed JW-KL scenarios for
%the $\theta$, providing a natural explanation of its
%narrow width~\cite{jm,carlson2}. 
The overlaps, $\langle NM_k\vert P_k\rangle$, depend on the structures 
of the hyperfine eigenstates and the basic spatial overlaps, 
\begin{equation}
x_k\, =\, \langle N_{123}M_{45}\vert [31]_L,k\rangle\ ,
\label{spatialoverlap}\end{equation}
with $k=SS, SA, AS$. $x_{AS}$ is identically zero as a result of the symmetry 
of the quark model $N$ spatial wavefunction. In addition, 
for the phase conventions employed here, $\vert [31]_L, SS\rangle$ 
transforms as $P_{23}\vert [31]_L, SS\rangle\, =\, 
{\frac{1}{\sqrt{2}}}\vert [31]_L, SA\rangle
\, -\, {\frac{1}{\sqrt{2}}}\vert [31]_L, AS\rangle$
under the action of the adjacent permutation $P_{23}$, leading to 
$x_{SS}={\frac{1}{\sqrt{2}}}x_{SA}$.
All overlaps can thus be written as a numerical coefficient
times the single common spatial overlap factor $x_{SA}$. The latter cancels
in forming ratios. Columns 4 and 5 of the tables contain,
for each excited pentaquark state $P^*$, 
the squares of the ratios $g_P$ and $g_{V^*}$,
defined by
\begin{eqnarray}
&&g_P\, =\, \langle NP_H\vert P^*\rangle 
/ \langle NP_H\vert P_{gnd}\rangle\nonumber\\
&&g_{V^*}\, =\, \langle NV^*_H\vert P^*\rangle
/ \langle NP_H\vert P_{gnd}\rangle \ ,
\label{gdefns}\end{eqnarray}
%$\langle NP_H\vert P^*\rangle$ and $\langle NV^*_H\vert P^*\rangle$
%to $\langle NP_H\vert P_{gnd}\rangle$, 
with $P_{gnd}$ the corresponding $(I,J_q)=(0,1/2)$ ground state.
% $g_P=0$ for $(I,J_q=3/2)$ states.

\begin{table}
\caption{\label{tablegb}
Low-lying positive parity excitations of the $\theta_{c,b}$ in the
GB model. All notation is as described in the text.}
\vskip .15in\noindent
\begin{ruledtabular}
\begin{tabular}{lcccc}
$\ (I,J_q)$&$\Delta\hat{E}$&$\Delta E^{est}$ (MeV)&$g_P^2$&$g_{V^*}^2$\\
\hline
(0,1/2)&0&0&1&3.00\\
(1,1/2)&4.50$\rightarrow$5.71&132$\rightarrow$167&2.24$\rightarrow$2.54
&0.75$\rightarrow$0.85\\
(1,3/2)&4.50$\rightarrow$5.71&132$\rightarrow$167&0
&1.27$\rightarrow$1.36\\
(0,1/2)&10.2$\rightarrow$14.5&299$\rightarrow$423&2.01$\rightarrow$2.07
&0.67$\rightarrow$0.69\\
(0,3/2)&10.2$\rightarrow$14.5&299$\rightarrow$423&0
&2.68$\rightarrow$2.75\\
\end{tabular}
\end{ruledtabular}
%\vspace*{-13pt}
\end{table}

%\vskip .2in\noindent
\begin{table}
\caption{\label{tablecmcb}
Low-lying positive parity excitations of the $\theta_{c,b}$ in the
CM model. All notation is as described in the text.}
%{\footnotesize
\begin{ruledtabular}
\begin{tabular}{llcccc}
Sector&$\ (I,J_q)$&$\Delta\hat{E}$&$\Delta E^{est}$ (MeV)&$g_P^2$&$g_{V^*}^2$\\
\hline
\hline
Charm&(0,1/2)&0&0&1&0.74$\rightarrow$2.22\\
\hline
&(0,1/2)&1.14$\rightarrow$1.20&84$\rightarrow$88&0.55$\rightarrow$1.87
&1.54$\rightarrow$2.32\\
&(1,1/2)&1.22$\rightarrow$1.47&89$\rightarrow$108&1.95$\rightarrow$3.41
&0.03$\rightarrow$0.35\\
&(0,3/2)&1.29$\rightarrow$1.56&94$\rightarrow$114&0
&1.60$\rightarrow$2.79\\
&(1,3/2)&1.61$\rightarrow$1.87&118$\rightarrow$137&0
&0.85$\rightarrow$1.52\\
&(1,1/2)&1.79$\rightarrow$2.07&131$\rightarrow$152&0.00$\rightarrow$0.14
&1.72$\rightarrow$2.72\\
\hline
&(0,1/2)&3.08$\rightarrow$3.20&226$\rightarrow$234&0.29$\rightarrow$0.32
&0.39$\rightarrow$0.92\\
&(1,1/2)&3.59$\rightarrow$3.78&263$\rightarrow$276&0.07$\rightarrow$0.10
&0.00$\rightarrow$0.03\\
&(1,3/2)&3.82$\rightarrow$4.10&280$\rightarrow$300&0
&0.09$\rightarrow$0.13\\
&(0,3/2)&3.84$\rightarrow$3.93&281$\rightarrow$289&0
&0.06$\rightarrow$0.21\\
&(0,1/2)&3.96$\rightarrow$4.08&290$\rightarrow$298&0.05$\rightarrow$0.15
&0.09$\rightarrow$0.14\\
\hline
\hline
%Bottom&$\ (I,J_q)$&$\Delta\hat{E}$&$\Delta E^{est}$&$g_P^2$&$g_{V^*}^2$\\
Bottom&(0,1/2)&0&0&1.00&1.87$\rightarrow$2.71\\
\hline
&(0,1/2)&1.16$\rightarrow$1.25&85$\rightarrow$92&1.54$\rightarrow$2.32
&0.88$\rightarrow$0.94\\
&(0,3/2)&1.26$\rightarrow$1.35&92$\rightarrow$99&0
&2.51$\rightarrow$3.21\\
&(1,1/2)&1.43$\rightarrow$1.55&105$\rightarrow$114&1.76$\rightarrow$3.65
&0.20$\rightarrow$0.76\\
&(1,3/2)&1.58$\rightarrow$1.66&116$\rightarrow$122&0
&1.36$\rightarrow$1.76\\
&(1,1/2)&1.77$\rightarrow$1.99&130$\rightarrow$146&0.05$\rightarrow$0.46
&2.53$\rightarrow$2.76\\
\hline
&(0,1/2)&3.06$\rightarrow$3.12&224$\rightarrow$229&0.32$\rightarrow$0.39
&0.79$\rightarrow$1.10\\
&(1,1/2)&3.66$\rightarrow$3.88&268$\rightarrow$284&0.08$\rightarrow$0.13
&0.02$\rightarrow$0.04\\
&(1,3/2)&3.79$\rightarrow$3.98&278$\rightarrow$292&0
&0.12$\rightarrow$0.15\\
&(0,3/2)&3.91$\rightarrow$3.94&286$\rightarrow$289&0
&0.14$\rightarrow$0.25\\
&(0,1/2)&3.93$\rightarrow$3.99&288$\rightarrow$292&0.11$\rightarrow$0.19
&0.07$\rightarrow$0.11\\
\end{tabular}
\end{ruledtabular}
\end{table}

Certain general features of the results are evident
from the tables. First, for both models, two groups of excited states
exist, one with ``low'' (less than $\sim 160$
MeV) and one with ``high'' (comparable to, or slightly
less than $m_\Delta -m_N$) excitation energies. 
Second, the lowest of the excitation energies
is significantly smaller in the CM than in the GB model,
$\Delta E^{est}\simeq 85-90$ MeV versus $\simeq 130-160$ MeV. Third, the
spectrum of excitations is far denser for the CM model,
which has 10 excited states within $\sim m_\Delta -m_N$ of 
the ground state, compared to only 4 for the GB model.
(Even more striking, for the CM model,
the 5 ``low'' excitations and 5 ``high'' excitations
each lie in intervals of size $\sim 50-60$ MeV; even if
some rescaling of the splitting estimates is required, two regions
with a very dense spectrum of states are thus predicted.)
Fourth, for the GB model both of the ``low''
excitations have $I=1$, whereas low-lying excitations with
both $I=1$ and $I=0$ exist for the CM model. 
Fifth, in both models, the ``low'' excitations
have one or both of their overlaps to $NP_H$ or $NV_H^*$
comparable to, or larger than, the ground state overlap to $NP_H$.
As such, if one of the states is experimentally detectable, the
others should be as well. Finally, while the ``high'' excitations
in the GB model also have overlaps comparable to that of the ground state,
those in the CM model have strongly suppressed overlaps to
both $NP_H$ and $NV_H^*$, with the exception of the lowest
of these states, for which the $NV_H^*$ overlap is comparable
to the ground state $NP_H$ overlap for some range of the
input parameter values. The remainder of the ``high'' excitations
in the CM model should thus decay preferentially
to multiparticle final states, making them more challenging
to identify experimentally. Note that the range of overlap
values is greater for the charm than bottom system in the CM model, 
reflecting the sensitivity of the overlaps to mixing effects, 
which are greater for lighter $m_{\bar{Q}}$. 
The excitation energies are typically
much less sensitive, especially so for the ``low'' group of states.

\section{Conclusions}
We have seen that rather low-lying spin-isospin excitations
are expected in the heavy $P=+$ pentaquark sector,
in both the GB and CM models. The number of such excitations
is especially large for the CM model. A spectrum of excitations richer
than in the ordinary baryon sector is generic to the quark model 
approach to pentaquark states since the number of Pauli-allowed states 
grows rapidly with the number of constituents. In the $[21]_L$, 
$\ell^3$ ordinary baryon sector, for example,
only three channels, with a single allowed
state each, are present, to be contrasted to the situation in the
pentaquark sector, where 4, 3, 1, 6, 5, 1, 2, 2, and 1 independent states
exist for the $[31]_L$, $\bar{Q}\ell^4$ $(I,J_q)=(0,1/2)$
$(0,3/2)$, $(0,5/2)$, $(1,1/2)$, $(1,3/2)$, $(1,5/2)$, 
$(2,1/2)$, $(2,3/2)$ and $(2,5/2)$ channels, respectively.
Spin-orbit partners, which are expected to lie rather 
nearby in the CM model~\cite{cdls}, will make the model spectra even denser.

The pattern of these low-lying
excitations has, in addition, been shown to be very different 
for the two models. Not only is the number of states within $\sim 300$ MeV
of the $(I,J_q)=(0,1/2)$ ground state much larger for the CM model,
but also the minimum excitation energy and pattern of quantum
numbers of the ``low'' group of excitations is significantly different
from that of the GB model. Since these states all have an
overlap to either $NP_H$ or $NV_H^*$ comparable to
or larger than that of the ground state to $NP_H$,
the presence of only $I=0$ states in the ``low'' excitation
region for the GB model, as well as the presence of such excitation
with both $I=0$ and $I=1$ in the CM model, should be experimentally
detectable by studying two-body decay modes. Experimental results
should thus allow one
to rule out at least one, and perhaps both, of the models.

\begin{acknowledgments}
The ongoing support of the Natural Sciences and Engineering Council
of Canada is gratefully acknowledged.
\end{acknowledgments}



\begin{thebibliography}{99}
\bibitem{earlyheavy} C. Gignoux, B. Silvestre-Brac and J.M. Richard,
Phys. Lett. {\bf B193} (1987) 323; H.J. Lipkin, Phys. Lett. {\bf B195}
(1987) 484; J. Leandri and B. Silvestre-Brac, Phys. Rev. {\bf D40}
(1989) 2340
%%CITATION = PHLTA,B193,323;%%
%%CITATION = PHLTA,B195,484;%%
%%CITATION = PHRVA,D40,2340;%%
\bibitem{nextheavy} S. Fleck, C. Gignoux, J.M. Richard and B. Silvestre-Brac,
Phys. Lett. {\bf B220} (1989) 616; S. Zouzou and J.M. Richard, Few-Body
Syst. {\bf 16} (1994) 1
%%CITATION = PHLTA,B193,323;%%
%%CITATION = HEP-PH 9309303;%%
\bibitem{glozman96}L. Ya. Glozman and D. O. Riska, Phys. Rep. {\bf 268}
(1996) 263
\bibitem{gbposparheavy} F. Stancu, Phys. Rev. {\bf D58} (1998) 111501
%%CITATION = HEP-PH 9803442;%%
\bibitem{gbnegparheavy} M. Genovese {\it et al.},
Phys. Lett. {\bf B425} (1998) 171
%%CITATION = HEP-PH 9712452;%%
\bibitem{e791heavy} E.M. Aitala {\it et al.}, (The E791 Collaboration),
Phys. Lett. {\bf B448} (1999) 303
%%CITATION = PHLTA,B448,303;%%
\bibitem{thetaexp}T. Nakano, {\it et al.}, (The LEPS Collaboration),
Phys. Rev. Lett. {\bf 91} (2003) 012002; 
V. Barmin, {\it et al.}, (The DIANA Collaboration), 
Phys. At. Nucl. {\bf 66} (2003) 1715;
S. Stepanyan, {\it et al.}, (The CLAS Collaboration), 
Phys. Rev. Lett. {\bf 91} (2003) 252001;
J. Barth, {\it et al.}, (The SAPHIR Collaboration),
Phys. Lett. {\bf B572} (2003) 127;
A.E. Asratyan, A.G. Dolgolenko and M.A. Kubantsev, Phys. Atom.
Nucl. {\bf 67} (2004) 682;
V. Kubarovsky {\it et al.}, (The CLAS Collaboration),
Phys. Rev. Lett. {\bf 92} (2004) 032001 (Erratum {\it ibid.}
{\bf 92} (2004) 049902);
H.G. Juengst {\it et al.}, (The CLAS Collaboration), nucl-ex/0312019; 
A. Airapetian {\it et al.}, (The HERMES Collaboration), 
Phys. Lett. {\bf B585} (2004) 213;
A. Aleev {\it et al.}, (The SVD-2 Collaboration), hep-ex/0401024;
M. Abdel-Bary {\it et al.}, (The COSY-TOF Collaboration), hep-ex/0403011;
P.Z. Aslanyan, V.N. Emelyanenko and G.G. Rikhkvitzkaya, hep-ex/0403044;
S. Chekanov {\it et al.}, (The ZEUS Collaboration), Phys. Lett.
{\bf B591} (2004) 7;
T. Nakano, presentation at NSTAR 2004, Grenoble, France, Mar. 24-27, 2004
(http://lpsc.in2p3.fr/congres/nstar2004/talks/nakano.pdf)
%%CITATION = HEP-EX 0312044;%%
%%CITATION = HEP-EX 0301020;%%
%%CITATION = HEP-EX 0304040;%%
%%CITATION = HEP-EX 0307018;%%
%%CITATION = HEP-EX 0307083;%%
%%CITATION = HEP-EX 0309042;%%
%%CITATION = HEP-EX 0311046;%%
%%CITATION = NUCL-EX 0312019;%%
%%CITATION = HEP-EX 0312044;%%
%%CITATION = HEP-EX 0401024;%%
%%CITATION = HEP-EX 0403011;%%
%%CITATION = HEP-EX 0403044;%%
%%CITATION = HEP-EX 0403051;%%
\bibitem{xi32}C. Alt {\it et al.}, (The NA49 Collaboration), 
Phys. Rev. Lett. {\bf 92} (2004) 042003;
K.T. Knopfle {\it et al.}, (The HERA-B Collaboration),
hep-ex/0403020; H.G. Fischer and S. Wenig, hep-ex/0401014
%%CITATION = HEP-EX 0310014;%% 
%%CITATION = HEP-EX 0403020;%% 
%%CITATION = HEP-EX 0401014;%%
\bibitem{H1} A. Aktas {\it et al.}, (The H1 Collaboration), 
hep-ex/0403017 
\bibitem{zeus} K. Lipka {\it et al.}, 
(for the H1 and ZEUS Collaborations), hep-ex/0405051
%%CITATION = HEP-EX 0401014;%%
%%CITATION = HEP-EX 0405051;%%
\bibitem{jw} R. Jaffe and F. Wilczek, Phys. Rev. Lett.
{\bf 91} (2003) 232003
%%CITATION = HEP-PH 0307341;%%
\bibitem{klth} M. Karliner and H.J. Lipkin, hep-ph/0307343
%%CITATION = HEP-PH 0307343;%%
\bibitem{huang} K. Cheung, Phys. Rev. {\bf D69} (2004) 094029;
P.Z. Huang {\it et al.}, hep-ph/0401191
%%CITATION = HEP-PH 0308176;%%
%%CITATION = HEP-PH 0401191;%%
\bibitem{rosner}J. Rosner, Phys. Rev. {\bf D69} (2004) 094014; 
X.G. He and X.Q. Li, hep-ph/0403191; T.W. Chiu and T.H. Hsieh, hep-ph/0404007;
H.C. Kim, S.H. Lee and Y.S. Oh, hep-ph/0404170
%%CITATION = HEP-PH 0312269;%%
%%CITATION = HEP-PH 0403191;%%
%%CITATION = HEP-PH 0404007;%%
%%CITATION = HEP-PH 0404170;%%
\bibitem{sww} I.W. Stewart, M.E. Wessling and M.B. Wise, Phys. Lett.
{\bf B590} (2004) 185
%%CITATION = HEP-PH 0402076;%%
\bibitem{csmthetac}B. Wu and B.Q. Ma, hep-ph/0402244;
M.A. Nowak, M. Praszalowicz, M. Sadzikowski and J. Wasliuk,
hep-ph/0403184
%%CITATION = HEP-PH 0402244;%%
%%CITATION = HEP-PH 0403184;%%
\bibitem{otherqm} H.Y. Cheng and C.K. Chua,
hep-ph/0403232; Y.R. Liu {\it et al.}, hep-ph/0404123; K. Cheung,
hep-ph/0405281; H.Y. Cheng and C.K. Chua, hep-ph/0406036
%%CITATION = HEP-PH 0403232;%%
%%CITATION = HEP-PH 0404123;%%
%%CITATION = HEP-PH 0405281;%%
%%CITATION = HEP-PH 0406036;%%
\bibitem{nussinov} S.~Nussinov, hep-ph/0307357
%%CITATION = HEP-PH 0307357;%%
\bibitem{kl} M. Karliner and H.J. Lipkin, Phys. Lett. {\bf B575}, 249 (2003)
and hep-ph/0307243
%%CITATION = HEP-PH 0402260;%%
\bibitem{earlycsm}A.V. Manohar, Nucl. Phys. {\bf B248} (1984) 19;
M. Chemtob, Nucl. Phys. {\bf B256} (1985) 600;
M. Praszalowicz in {\it Skyrmions and Anomalies},
M. Jezabeck and M. Praszalowicz eds., World Scientific, 1987
(see also M.~Praszalowicz, Phys. Lett.{\bf  B575} (2003) 234)
%%CITATION = NUPHA,B248,19;%%
%%CITATION = NUPHA,B256,600;%%
%%CITATION = HEP-PH 0308114%%;
\bibitem{dpp} D. Diakonov, V. Petrov, and M. Polyakov, Z. Phys. {\bf A359}
(1997) 305
%%CITATION = HEP-PH 9703373;%%
\bibitem{csmothers} H. Weigel, Eur. Phys. J. {\bf A2} (1998) 391;
H. Weigel, Proc. Intersections of Particle and
Nuclear Physics, hep-ph/0006191;
H .Walliser and V.B. Kopeliovich, J. Exp. Theor. Phys. {\bf 97} (2003)
433; V.B. Kopeliovich, hep-ph/0310071;
D. Borisyuk, M. Faber, A. Kobushkin, hep-ph/0307370 and
hep-ph/0312213;
D. Diakonov and V. Petrov, Phys. Rev. {\bf D69} (2004) 094011;
B. Wu and B.Q. Ma, Phys. Rev. {\bf D69} (2004) 077501 and
Phys. Lett. {\bf B586} (2004) 62
%%CITATION = HEP-PH 0310212;%%
%%CITATION = HEP-PH 0307370;%%
%%CITATION = HEP-PH 9804260;%%
%%CITATION = HEP-PH 0006191;%%
%%CITATION = HEP-PH 0304058;%%
%%CITATION = HEP-PH 0310071;%%
%%CITATION = HEP-PH 0307370;%%
%%CITATION = HEP-PH 0310212;%%
%%CITATION = HEP-PH 0312041;%%
%%CITATION = HEP-PH 0312213;%%
%%CITATION = HEP-PH 0312326;%%
\bibitem{ekp} J. Ellis, M. Karliner and M. Praszalowicz, JHEP {\bf 0405}
(2004) 002
%%CITATION = HEP-PH 0401127;%%
\bibitem{kmdib} K. Maltman, Phys. Lett. {\bf B291} (1992) 371
%%CITATION = PHLTA,B291,373;%%
\bibitem{closezhao} F.E. Close and Q. Zhao, Phys. Lett. {\bf B590} (2004)
176
%%CITATION = HEP-PH 0403159;%%
\bibitem{jm} B. Jennings and K. Maltman, Phys. Rev. {\bf D69} (2004) 094020 
%%CITATION = HEP-PH 0308286;%%
\bibitem{cdls} J.J. Dudek and F.E. Close, Phys. Lett. {\bf B583} (2004) 278
%%CITATION = HEP-PH 0311258;%%

%\bibitem{carlson2} C.E. Carlson {\it et al.}, hep-ph/0312325.
%\bibitem{couplingwidth} N. Auerbach and V. Zelevinsky, nucl-th/0310029;
%M. Karliner and H. Lipkin, Phys. Lett. {\bf B586}, 303 (2004).



\end{thebibliography}
\end{document}